\documentclass[11pt,a4paper]{article}
\pdfoutput=1
\usepackage{graphicx}
\usepackage{jcappub}

\usepackage{bm}
\usepackage{amsfonts}
\usepackage{subfigure}

\newcommand{\ns}{\Delta N_{\rm {eff}}}
\newcommand{\Neff}{N_{\rm {eff}}}

\newcommand{\be}{\begin{equation}}
\newcommand{\ee}{\end{equation}}
\newcommand{\bea}{\begin{eqnarray}}
\newcommand{\eea}{\end{eqnarray}}

\begin{document}

%%%%%%%%%%%%%%%%%%%%%%%%%%%%%%%%%%%%%%%%%%%%%%%%%%%%%%%%%%%%%%%%%%%%%%
% Frontpage %%%%%%%%%%%%%%%%%%%%%%%%%%%%%%%%%%%%%%%%%%%%%%%%%%%%%%%%%%
%%%%%%%%%%%%%%%%%%%%%%%%%%%%%%%%%%%%%%%%%%%%%%%%%%%%%%%%%%%%%%%%%%%%%%

%\subheader{\hfill Preprint ....}

\title{Light sterile neutrinos after BICEP-2}

\author[a]{Maria Archidiacono}
\author[b]{Nicolao Fornengo}
\author[b]{Stefano Gariazzo}
\author[b]{Carlo Giunti}
\author[a,c]{Steen Hannestad}
\author[d]{Marco Laveder}

\affiliation[a]{Department of Physics and Astronomy, University of Aarhus\\
 DK-8000 Aarhus C, Denmark}

\affiliation[b]{Department of Physics, University of Torino and INFN, \\
Via P. Giuria 1, I--10125 Torino, Italy}

\affiliation[c]{Aarhus Institute of Advanced Studies,\\
 Aarhus University, DK-8000 Aarhus C, Denmark}

\affiliation[d]{Dipartimento di Fisica e Astronomia ``G. Galilei'', Universit\`a di Padova, and INFN, Sezione di Padova, \\
Via F. Marzolo 8, I--35131 Padova, Italy}

\emailAdd{archi@phys.au.dk}
\emailAdd{gariazzo@to.infn.it}

\abstract{The recent discovery of B-modes in the polarization pattern of the Cosmic Microwave Background by the BICEP2 experiment has important implications for neutrino physics. We revisit cosmological bounds on light sterile neutrinos and show that they are compatible with all current cosmological data provided that the mass is relatively low. Using CMB data, including BICEP-2, we find an upper bound of $m_s < 0.85$ eV ($2\sigma$ Confidence Level). This bound is strengthened to 0.48 eV when HST measurements of $H_0$ are included. However, the inclusion of SZ cluster data from the Planck mission and weak gravitational measurements from the CFHTLenS project favours a non-zero sterile neutrino mass of $0.44^{+0.11}_{-0.16}$ eV. 
Short baseline neutrino oscillations, on the other hand, indicate a new mass state around 1.2 eV. This mass is highly incompatible with cosmological data if the sterile neutrino is fully thermalised ($\Delta \chi^2>10$). However, if the sterile neutrino only partly thermalises it can be compatible with all current data, both cosmological and terrestrial.}

\maketitle

\section{Introduction}

Over the past few years cosmology has established itself as one of the primary laboratories for neutrino physics. In particular, observations of the Cosmic Microwave Background and Large Scale Structure have severely constrained parameters such as the absolute neutrino mass and the cosmic energy density in neutrinos (see e.g.\ \cite{Ade:2013zuv}). These two parameters are also of significant interest in the context of eV-mass sterile neutrinos currently hinted at by short baseline neutrino oscillation experiments.
At the same time neutrino oscillation experiments seem to point to the existence of at least one additional mass state around 1 eV with significant mixing with the active sector. Even though this mass state is mainly sterile the mixing leads to almost complete thermalisation in the early universe (see e.g.\ 
\cite{Hannestad:2012ky, Melchiorri:2008gq}) and the additional mass state effectively affects structure formation in the same way as a 1 eV active neutrino. Such a high mass has seemed at odds with cosmological data \cite{Hamann:2011ge}, and has led to a number of attempts to reconcile the existence of eV sterile neutrinos with cosmology. Examples include modifications to the background potential due to new interactions in the sterile sector \cite{Hannestad:2013ana,Dasgupta:2013zpn,Bringmann:2013vra,Ko:2014bka,Mirizzi:2012we,Saviano:2013ktj} or modifications to the cosmic expansion rate at the time where sterile neutrinos are produced \cite{Rehagen:2014vna}. 

However, the very recent publication of new data from the BICEP2 experiment \cite{Ade:2014xna} has indicated a high tensor to scalar ratio, and this in turn significantly modifies constraints on neutrino related parameters. Here we investigate how constraints on eV mass sterile neutrinos are influenced by the new BICEP2 discovery, and demonstrate that eV mass sterile neutrinos are not significantly constrained by current cosmological data.

Section 2 contains a discussion of the cosmological parameter estimation and Section 3 a short summary of our SBL likelihood analysis. In Section 4 we present the results of the joint analysis and finally Section 5 contains a thorough discussion of our results.

\section{The cosmological analysis}

The setup under investigation here is a model in which the neutrino sector is described by 3 massless or almost massless active species, as well as one additional sterile species characterised by a temperature, $T_s$. We thus assume that the sterile neutrino has a thermal distribution. Although this almost certainly does not happen unless the sterile species was fully thermalised, it is a more than adequate approximation given the precision of current cosmological data. From the temperature and the mass the contribution to the current matter density is given by
\begin{equation}
\Omega_s h^2 = \frac{(T_s/T_\nu)^3 m_s}{94 \, {\rm eV}},
\end{equation}
where $T_\nu$ is the temperature of the active species and $m_s$ the mass of the additional sterile neutrino.
Likewise the contribution to the relativistic energy density in the early universe is given by
\begin{equation}
\rho_s = (T_s/T_\nu)^4 \rho_\nu.
\end{equation}
Most studies work with an effective number of neutrino species, defined by $\ns = (T_s/T_\nu)^4$, and we shall also use this parameter in order for our results to be easily comparable other studies. In terms of $\ns$ we have
\begin{eqnarray}
\Omega_s h^2 & = & \frac{\ns^{3/4} m_s}{94 \, {\rm eV}}, \\
\rho_s & = & \ns \rho_\nu.
\end{eqnarray}
Our cosmological model is a flat $\Lambda$CDM+$r_{0.002}$+$\nu_s$ model with a total of 
nine parameters
\begin{equation}\label{eq:model}
{\bm \theta} = \{\omega_{\rm cdm},\omega_{\rm b},\theta_{\rm s},\tau,
\ln(10^{10}A_{s}),n_{s},r_{0.002},m_s,\ns\}.
\end{equation}
Here, $\omega_{\rm cdm} \equiv \Omega_{\rm cdm} h^2$ and $\omega_{\rm b}
\equiv \Omega_{\rm b} h^2$ are the present-day physical CDM and baryon densities
respectively, $\theta_{\rm s}$ the angular the sound horizon, $\tau$  the optical
depth to reionisation, and $\ln(10^{10}A_{s})$ and $n_s$ denote respectively the amplitude and spectral index
of the initial scalar fluctuations. $r$ is the tensor to scalar ratio at the pivot scale of $0.002 \, {\rm Mpc}^{-1}$. We assume a flat prior on all of the cosmological parameters but the $m_s$; in the case of the physical mass of the additional sterile neutrino the posterior obtained through the analysis of neutrino oscillations data (see Sec.~\ref{sec:neutrinooscillationdata}) is applied as a prior on the cosmological parameter $m_s$.

The bayesian analysis is performed through the Monte Carlo Markov Chains package \texttt{CosmoMC} \cite{Lewis:2002ah}. The calculation of the theoretical observables is done through the Boltzman equations solver \texttt{CAMB} \cite{Lewis:1999bs} (Code for Anisotropies in the Microwave Background).

\subsection{Data sets}

This paper is aimed at testing the consistency between the latest cosmological data and the neutrino oscillation data (hereafter SBL). The former consist of CMB data, Large Scale Structure, Hubble constant $H_0$, CFHTLenS and Planck Sunyaev Zel'Dovich.

{\it CMB ---}
The primary cosmological observable in the early universe is the Cosmic Microwave Background. Therefore our basic data sets are: the temperature fluctuations power spectra provided by the Planck satellite \cite{Planck:2013kta} up to $\ell=2479$ and by Atacama Cosmlogy Telescope \cite{Dunkley:2013vu} and South Pole Telescope \cite{Story:2012wx} (hereafter high-$\ell$) whose likelihoods cover the high multipole range,  $500<\ell<3500$ and $650<\ell<3000$, respectively. Concerning polarization we include the data of the Wilkinson Microwave Anisotropy Probe nine year data release (hereafter WP) and the newly released B-modes autocorrelation power spectrum of the BICEP2 experiment, either using all of the nine channels ($20<\ell<340$), or only the first five data points ($\ell<200$), as in the BICEP2 paper \cite{Ade:2014xna}.

{\it Large Scale Structure (LSS) ---}
The information on Large Scale Structure is extracted from the WiggleZ Dark Energy Survey \cite{Parkinson:2012vd}, which measures the matter power spectrum at four different redshifts $z=0.22$, $z=0.41$, $z=0.60$ and $z=0.78$.

{\it $H_0$ ---}
The cosmological observable in the local universe consists of the distance measurements of the Cepheids obtained with the Hubble Space Telescope. These measurements provide a precise determination of the Hubble constant \cite{Riess:2011yx}, which acts as a prior on the derived cosmological parameter $H_0$.

{\it CFHTLenS ---}
The Canada-France Hawaii Telescope Lensing Survey (CFHTLenS) \cite{Kilbinger:2012qz,Heymans:2013fya} determines the 2D cosmic shear correlation function through the measurements of redshifts and shapes of 4.2 million galaxies spanning the range $0.2<z<1.3$. The weak gravitational lensing signal extracted from these measurements constrains a combination of the total matter density and the standard deviation of the amplitude of the matter density fluctuations on a sphere of radius $8{\rm h}^{-1}{\rm Mpc}$: $\sigma_8(\Omega_m/0.27)^{0.46}=0.774\pm0.040$. This result is included in our analysis, contributing as an additional $\chi^2$.

{\it PSZ ---}
The Planck Sunayev Zel'Dovich catalogue \cite{Ade:2013lmv} contains 189 galaxy clusters identified through the Sunayev Zel'Dovich effect. The number counts allows to compute the cluster mass function, which is related to a combination of $\Omega_m$ and $\sigma_8$: $\sigma_8(\Omega_m/0.27)^{0.3}=0.782\pm0.010$. This result is incorporated in our analysis following the same prescription used for CFHTLenS.

\section{Neutrino oscillation data}
\label{sec:neutrinooscillationdata}

Sterile neutrinos are new particles beyond the Standard Model which can mix with the standard active flavor neutrinos
$\nu_{e}$,
$\nu_{\mu}$,
$\nu_{\tau}$
(see \cite{Bilenky:1998dt,GonzalezGarcia:2007ib,Abazajian:2012ys}).
In the standard three-neutrino mixing paradigm
the three active flavor neutrinos
are unitary linear combinations of three massive neutrinos
$\nu_{1}$,
$\nu_{2}$,
$\nu_{3}$
with respective masses
$m_{1}$,
$m_{2}$,
$m_{3}$.
The squared mass differences
$
\Delta{m}^{2}_{21}
\simeq
8 \times 10^{-5} \, \text{eV}^{2}
$
and
$
\Delta{m}^{2}_{31}
\simeq
\Delta{m}^{2}_{32}
\simeq
2 \times 10^{-3} \, \text{eV}^{2}
$
generate the neutrino oscillations which have been observed
in many solar, atmospheric and long-baseline experiments
(see \cite{GonzalezGarcia:2012sz,Bellini:2013wra,NuFIT-2013,Capozzi:2013csa}).
However, the standard three-neutrino mixing paradigm
cannot explain the indications in favor of short-baseline
neutrino oscillations
found in the LSND experiment
\cite{Aguilar:2001ty},
in Gallium experiments
\cite{Abdurashitov:2005tb,Laveder:2007zz,Giunti:2006bj,Acero:2007su,Giunti:2010zu}
and in reactor experiments
\cite{Mueller:2011nm,Mention:2011rk,Huber:2011wv}.
The results of these experiments can be explained by extending neutrino mixing
with the addition of one or more massive neutrinos
which generate squared-mass differences larger than about $1 \, \text{eV}^2$
\cite{Kopp:2011qd,Giunti:2011gz,Giunti:2011hn,Giunti:2011cp,Conrad:2012qt,Kopp:2013vaa,Giunti:2013aea}.
Since there are only three active flavor neutrinos,
the additional massive neutrinos must be mainly sterile.

In this paper we consider a 3+1 scheme in which the three standard massive neutrinos
$\nu_{1}$,
$\nu_{2}$,
$\nu_{3}$
are much lighter than 1 eV
and there is a new massive neutrino $\nu_{4}$
with a mass $m_{4} \sim 1 \, \text{eV}$.
In the flavor basis the three standard active flavor neutrinos
$\nu_{e}$,
$\nu_{\mu}$,
$\nu_{\tau}$
are maily composed by
$\nu_{1}$,
$\nu_{2}$,
$\nu_{3}$,
but they have a small component of $\nu_{4}$ in order to generate the observed short-baseline oscillations
through the squared-mass difference
$
\Delta{m}^{2}_{43}
\simeq
\Delta{m}^{2}_{42}
\simeq
\Delta{m}^{2}_{41}
\simeq
m_{4}^2
$.
In the flavor basis there is a sterile neutrino $\nu_{s}$ which is mainly composed of
the new heavy neutrino $\nu_4$.
Hence,
in the following we use the common notation $m_{s}=m_{4}$.

We perform a combined analysis of cosmological data and
short-baseline neutrino oscillation data
using the posterior distribution of
$m_s = m_4 \simeq \sqrt{\Delta{m}^2_{41}}$
obtained from the analysis of SBL data presented in Ref.~\cite{Giunti:2013aea}
as a prior in the \texttt{CosmoMC} analysis of cosmological data
\cite{Archidiacono:2012ri,Archidiacono:2013xxa,Gariazzo:2013gua}.
As shown in Tab.~3 of Ref.~\cite{Gariazzo:2013gua},
the best-fit value of $m_s$ obtained from short-baseline neutrino oscillation data
is 1.27 eV
and its 95.45\% probability range ($2\sigma$)
is between
0.97 and 1.42 eV.

\section{Results}

{\it Cosmological results ---}

\begin{figure}[ht]
\centering
\includegraphics[scale=0.4]{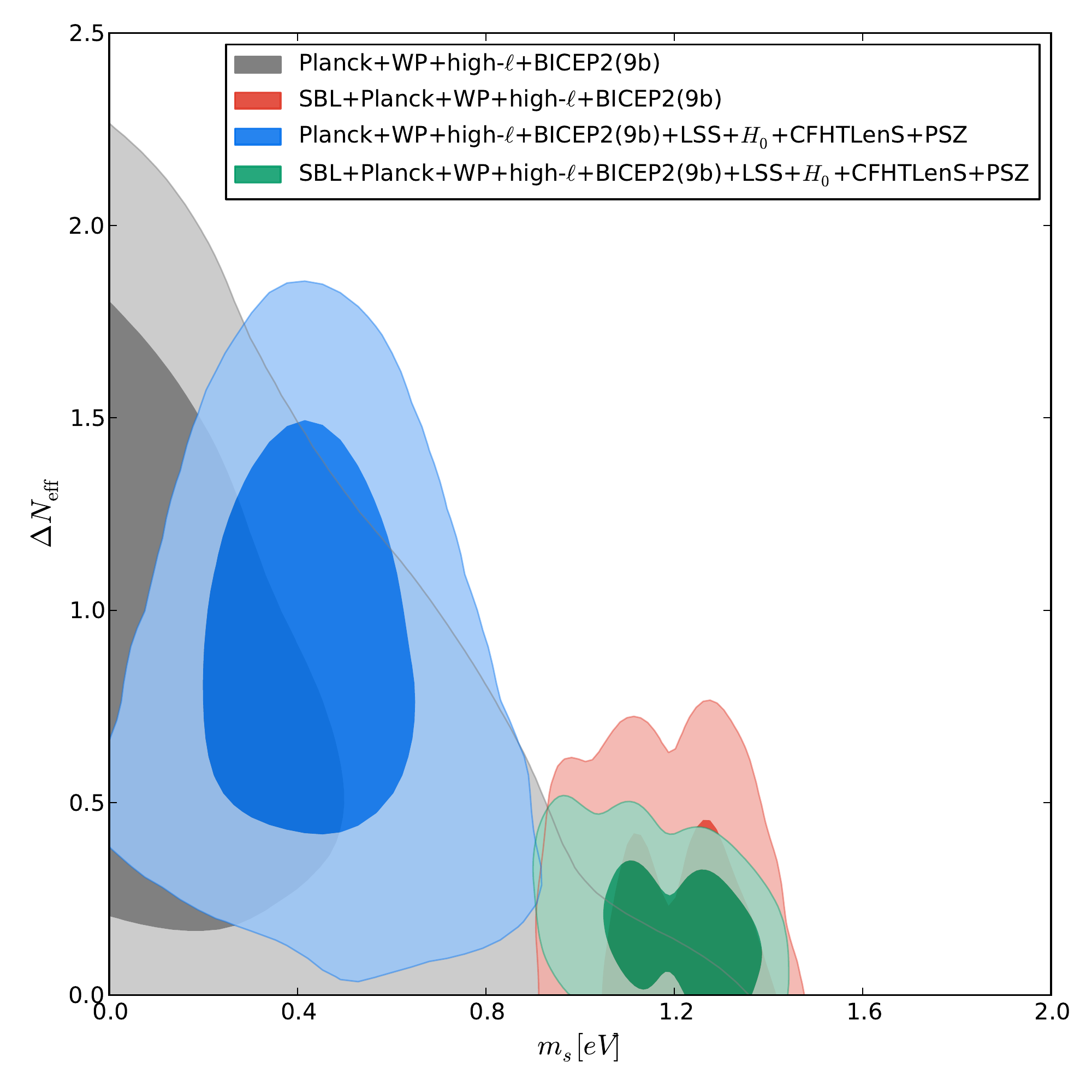} \\
\caption{$1\sigma$ and $2\sigma$ marginalized contours in the plane $(m_s,\ns)$. The banana shaped regions allowed by cosmology indicate a sub-eV mass and an excess in $\Neff$, while the inclusion of SBL data forces the mass around 1eV, moving the contours towards the warm dark matter limit, which implies a lower value of $\ns$ because of the strong correlation between the two parameters.}
\label{fig:ns_ms}
\end{figure}

An interesting question is how the addition of the new BICEP-2 measurement changes the preferred region in $(m_s,\ns)$ (see Fig. \ref{fig:ns_ms}) space.

\begin{figure}[ht]
\centering
\includegraphics[scale=0.3]{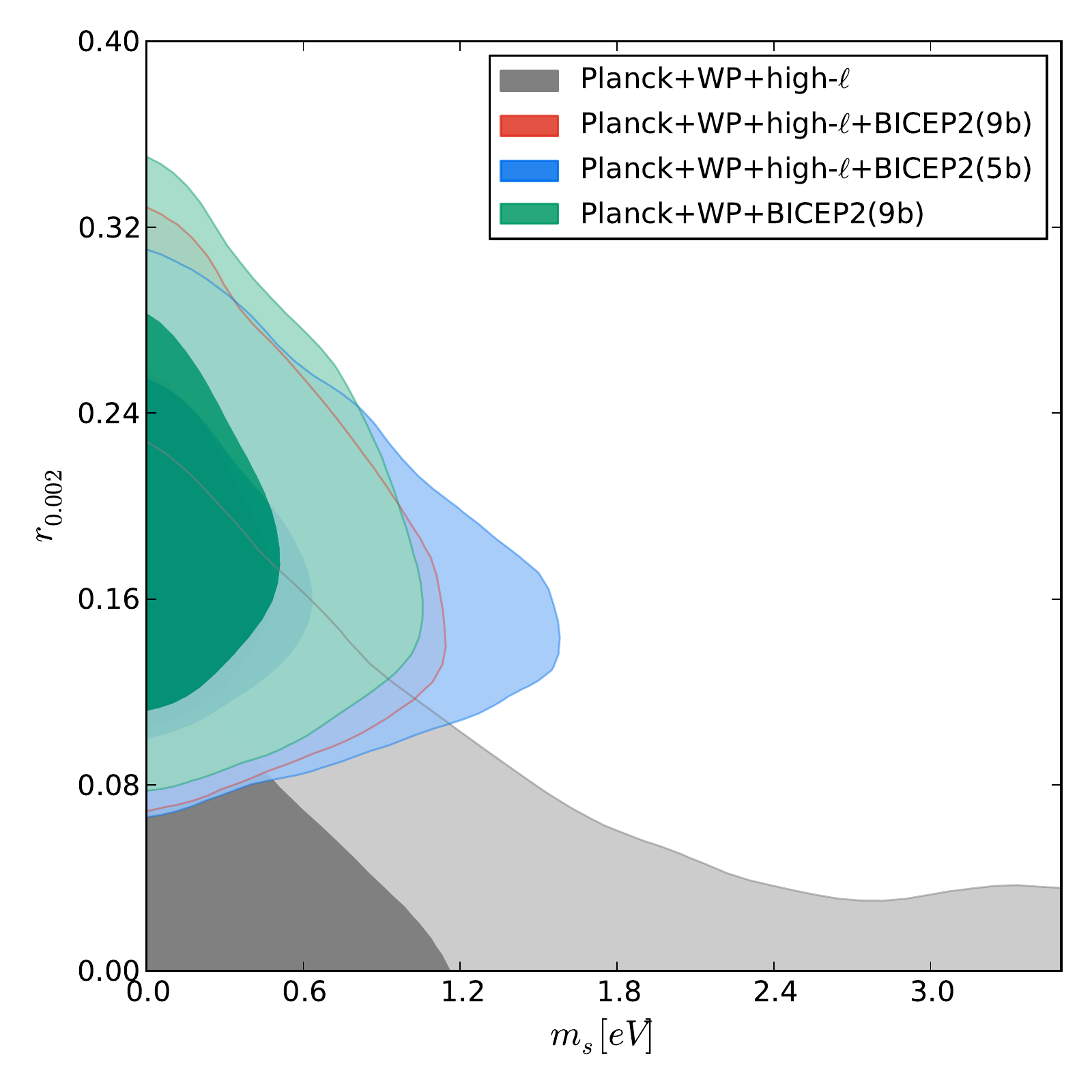} \includegraphics[scale=0.3]{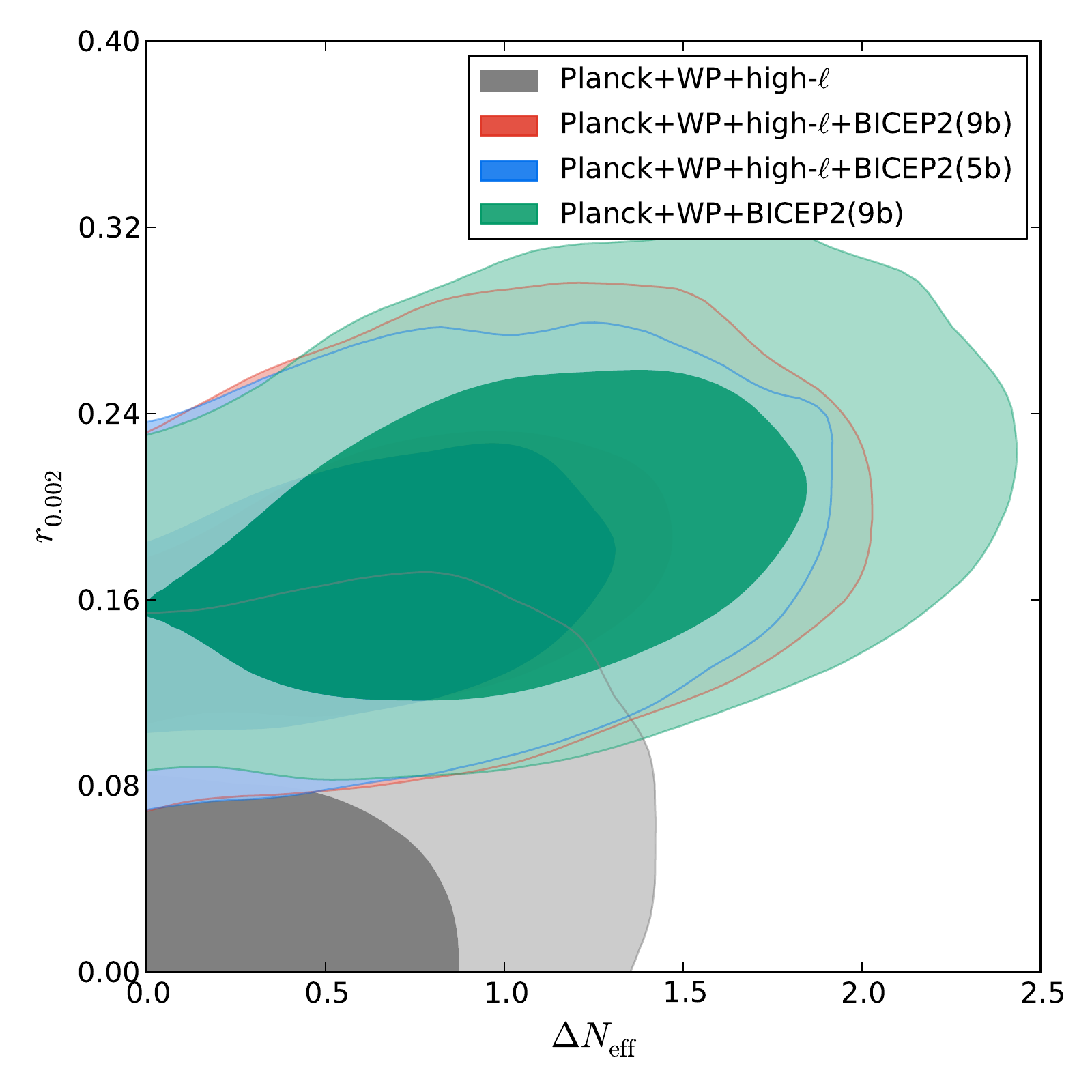} \includegraphics[scale=0.3]{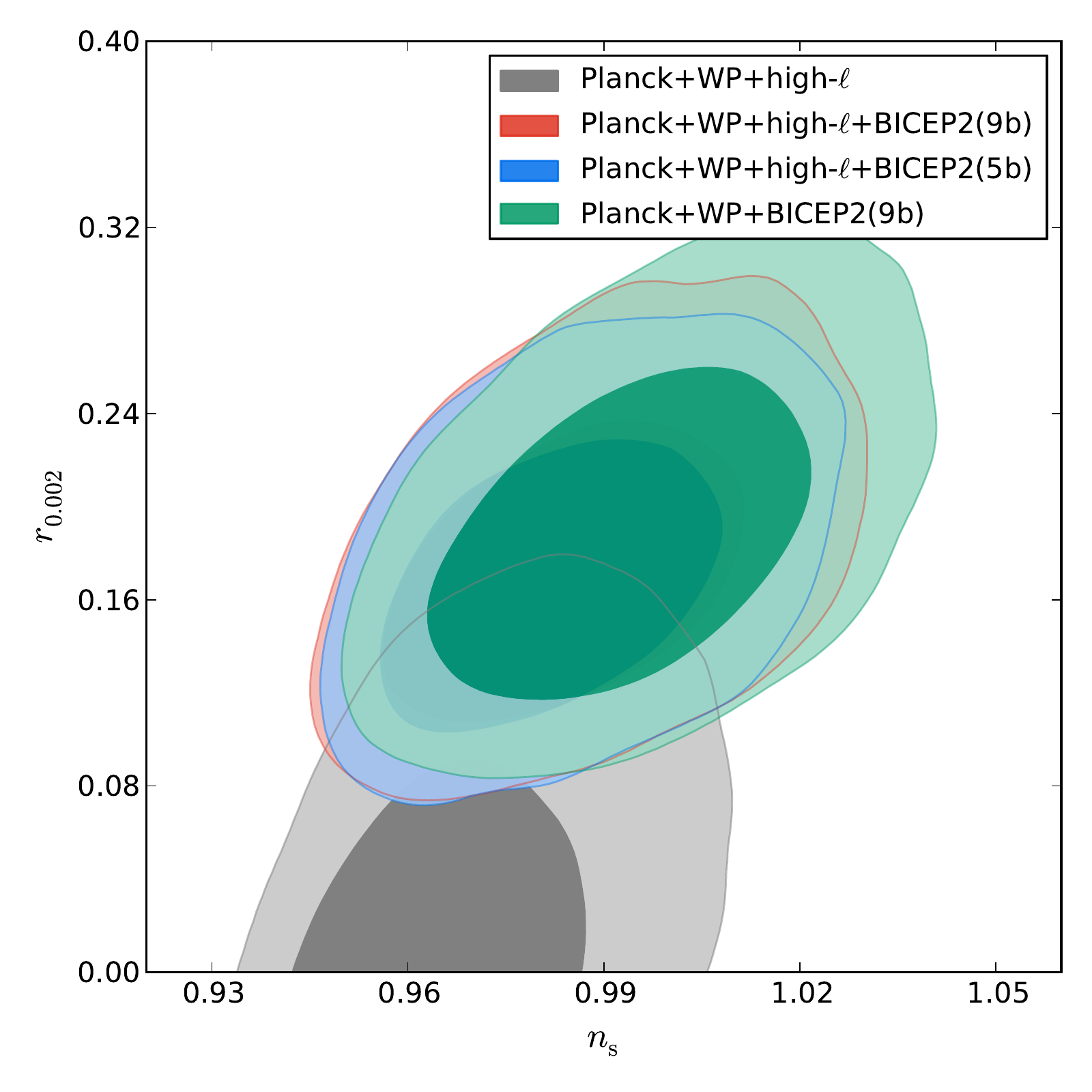} %\includegraphics[scale=0.25]{ns_ms.pdf}\\
\caption{$1\sigma$ and $2\sigma$ marginalized contours for different combinations of CMB data sets.}
\label{fig:cmbonly}
\end{figure}

Therefore, we first look at CMB data only, with and without BICEP-2 data included. The result of this analysis can be seen in Fig.~\ref{fig:cmbonly} and in Tab.~\ref{tab:cmbonly}. As can be seen in Fig.~\ref{fig:cmbonly} $m_s$ and $r$ are anti-correlated (this happens because $r$ adds power on large scales whereas $m_s$ subtracts power on intermediate and small scales). The inclusion of BICEP-2 data therefore tends to strengthen the bound on $m_s$ in order to keep constant the ratio between the small and large scales. Conversely, adding BICEP-2 data allows for higher values of $\Neff$ (this happens because $\Neff$ is strongly correlated with $n_s$ and the addition of tensors shifts the allowed $n_s$ up).
For the case of CMB data only, the addition of BICEP-2 data therefore strengthens the bound on $m_s$ slightly while allowing for a much higher $\Neff$. This is consistent with the analysis presented in \cite{Dvorkin:2014lea}
\footnote{Notice that here the notation is different: in Ref.~\cite{Dvorkin:2014lea} $m_s$ indicates the effective mass of the sterile neutrino, while our $m_s$ is the physical mass. A direct comparison of the numerical results is not possible due to volume effects in Bayesian marginalization. Here we just want to emphasize that, concerning the effect due to the inclusion of BICEP-2 data, both our results and those of Ref.~\cite{Dvorkin:2014lea} point towards tighter constraints on the additional massive component.}. 
When the inclusion of the BICEP2 data is restricted to the first five bins, the results concerning the basic cosmological parameters remain unchanged within $1\sigma$, whereas the bound on the mass becomes slightly weaker and, conversely, $\ns$ is tighter constrained. Finally if we remove the high multipole CMB data, the bound on the mass remains almost unchanged, while $\ns$ moves towards one additional fully thermalized sterile neutrino.

\begin{table*}[ht]
\begin{center}
\resizebox{1\textwidth}{!}{
\begin{tabular}{|l|c|c|c|c|}
\hline
\hline
 Parameters 
 &Planck+WP+high-$\ell$ &Planck+WP+high-$\ell$&Planck+WP+high-$\ell$
 &Planck+WP \\
 & &+BICEP2(9bins) &+BICEP2(5bins) & +BICEP2(9bins)\\

\hline
$\Omega_{\rm b} h^2$ 	
			& $0.02231^{+0.00032}_{-0.00040}\,^{+0.00078}_{-0.00072}$ & $0.02251^{+0.00039}_{-0.00046}\,^{+0.00087}_{-0.00078}$ & $0.02249^{+0.00035}_{-0.00045}\,^{+0.00084}_{-0.00078}$ & $0.02259^{+0.00040}_{-0.00050}\,^{+0.00094}_{-0.00082}$ \\

$\Omega_{\rm cdm} h^2$ 
			& $0.125^{+0.005}_{-0.007}\,^{+0.011}_{-0.010}$           & $0.129^{+0.006}_{-0.007}\,^{+0.013}_{-0.012}$           & $0.128^{+0.005}_{-0.008}\,^{+0.013}_{-0.012}$           & $0.132^{+0.007}_{-0.008}\,^{+0.015}_{-0.014}$           \\

$\theta_{\rm s}$ 
			& $1.0404^{+0.0009}_{-0.0008}\,^{+0.0016}_{-0.0017}$      & $1.0399^{+0.0009}_{-0.0009}\,^{+0.0017}_{-0.0017}$      & $1.0401^{+0.0009}_{-0.0009}\,^{+0.0018}_{-0.0017}$      & $1.0395^{+0.0009}_{-0.0009}\,^{+0.0019}_{-0.0018}$      \\

$\tau$ 
			& $0.094^{+0.013}_{-0.016}\,^{+0.031}_{-0.027}$           & $0.097^{+0.013}_{-0.016}\,^{+0.031}_{-0.027}$           & $0.096^{+0.013}_{-0.016}\,^{+0.030}_{-0.029}$           & $0.098^{+0.014}_{-0.017}\,^{+0.031}_{-0.031}$           \\

$n_{\rm s}$ 
			& $0.970^{+0.011}_{-0.018}\,^{+0.033}_{-0.027}$           & $0.986^{+0.016}_{-0.020}\,^{+0.035}_{-0.033}$           & $0.983^{+0.014}_{-0.020}\,^{+0.034}_{-0.031}$           & $0.995^{+0.017}_{-0.021}\,^{+0.038}_{-0.036}$           \\

$\log(10^{10} A_s)$ 
			& $3.106^{+0.029}_{-0.036}\,^{+0.068}_{-0.062}$           & $3.120^{+0.030}_{-0.037}\,^{+0.071}_{-0.061}$           & $3.167^{+0.047}_{-0.040}\,^{+0.080}_{-0.089}$           & $3.145^{+0.052}_{-0.046}\,^{+0.090}_{-0.098}$           \\

$r$ 
			& $<0.145$                                                & $0.177^{+0.036}_{-0.050}\,^{+0.093}_{-0.086}$           & $0.172^{+0.035}_{-0.048}\,^{+0.088}_{-0.082}$           & $0.192^{+0.040}_{-0.055}\,^{+0.101}_{-0.092}$           \\
\hline
$\ns$ 
			& $<1.18$                                                 & $0.82^{+0.40}_{-0.57};\,<1.66$                           & $0.73^{+0.31}_{-0.59};\,<1.56$                          & $1.08^{+0.49}_{-0.61};\,<2.03$                          \\

$m_s [\rm{eV}]$
			& $<2.17$                                                 & $<0.85$                                                 & $<1.15$                                                 & $<0.81$                                                 \\

\hline
\hline
\end{tabular}
}
\vspace{0.3cm}
\caption{Marginalized $1\sigma$ and $2\sigma$ confidence level limits for the cosmological parameters in various dataset combinations, given with respect to the mean value. Upper limit are given at $2\sigma$.}
\label{tab:cmbonly}
\end{center}
\end{table*}

Having established how constraints change from CMB data only we now proceed to study the influence of auxiliary cosmological data.

In Tab.~\ref{tab:cosmoresults} we report the marginalized mean values and the $1\sigma$ and $2\sigma$ errors on the cosmological parameters and on the neutrino parameters in the different combinations of data sets illustrated above, when SBL data are not included.

\begin{figure}[ht]
\centering
\includegraphics[scale=0.4]{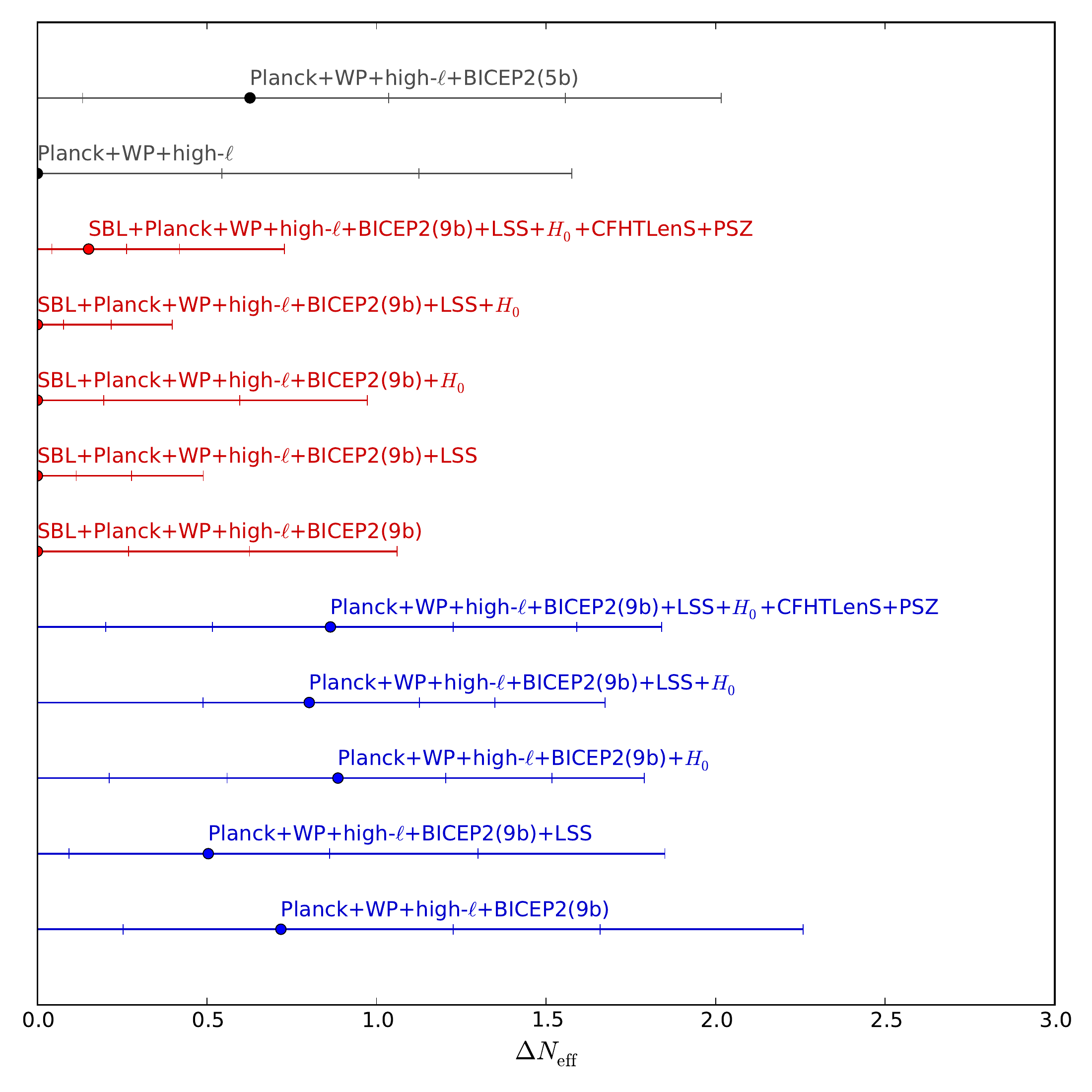} \\
\caption{$1\sigma$, $2\sigma$ and $3\sigma$ confidence level limits for $\ns$, for different dataset combinations. The circles indicate the mean value.}
\label{fig:ns}
\end{figure}

\begin{figure}[ht]
\centering
\includegraphics[scale=0.4]{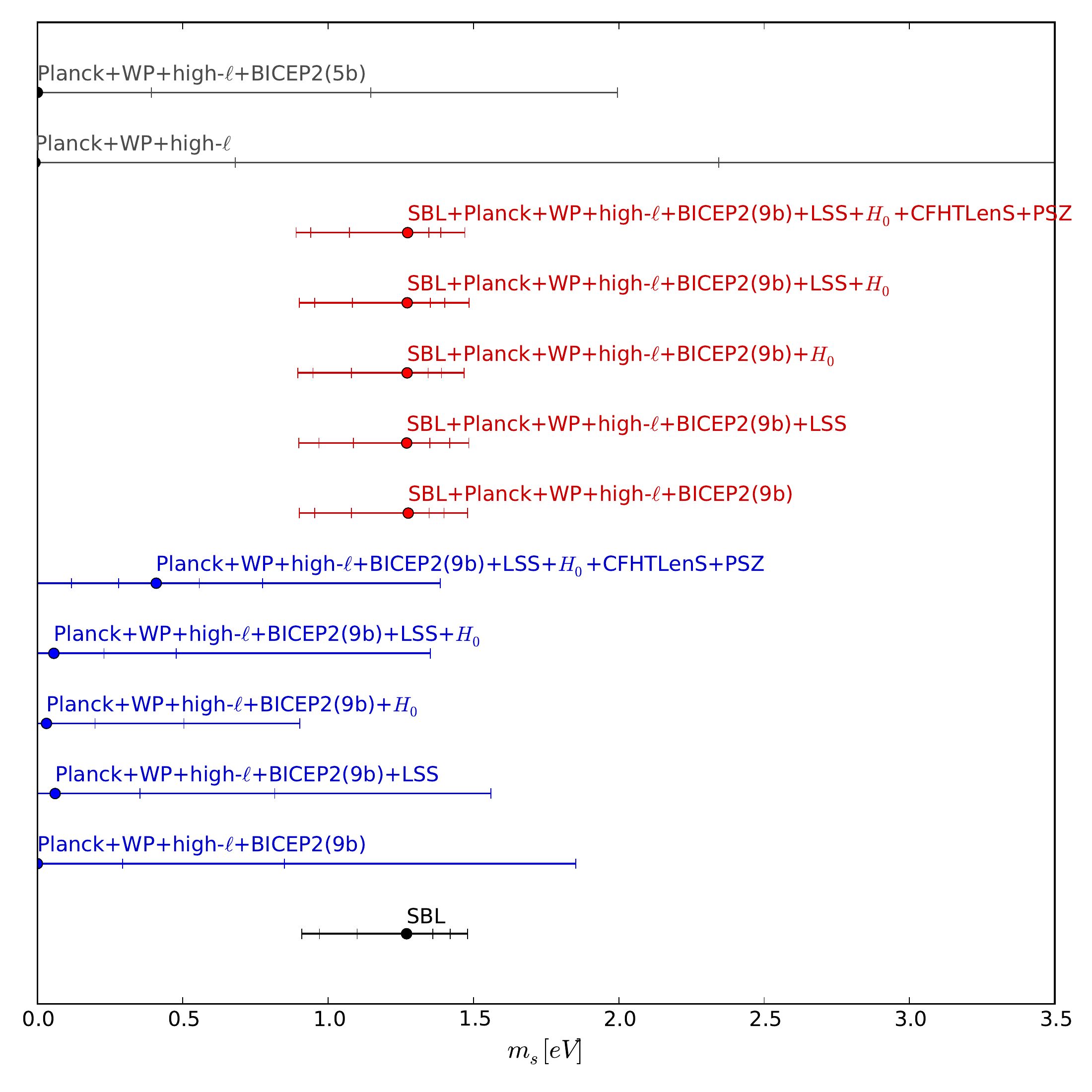} \\
\caption{$1\sigma$, $2\sigma$ and $3\sigma$ confidence level limits for $m_s$, for different dataset combinations. The circles indicate the mean value.}
\label{fig:ms}
\end{figure}

As was seen above, Planck CMB data provide a fairly stringent upper limit on the sterile neutrino mass, except for very low values of $\Neff$, i.e.\ in the warm dark matter limit. Conversely the preferred value of $\Neff$ is higher than 3, with 4 only being slightly disfavoured.
The inclusion of BICEP-2 data pushes the preferred $\Neff$ up, as has also been noted by other authors \cite{Giusarma:2014zza, Zhang:2014dxk, Dvorkin:2014lea}. However, since $m_s$ and $\Neff$ are anti-correlated this actually results in a tighter bound on the sterile neutrino mass from CMB only.

When we include LSS or $H_0$ data the picture remains qualitatively unchanged although, since $m_s$ and $H_0$ are anti-correlated, the addition of the HST $H_0$ data strengthens the upper bound on the sterile neutrino mass.
In Fig.~\ref{fig:ms} and Fig.~\ref{fig:ns} we can see how the error bars change for $m_s$ and $\ns$ respectively, with various dataset combinations.

However, the inclusion of lensing and cluster data leads to an important qualitative change the preferred range for $m_s$. Both data sets indicate a low value of $\sigma_8$. Given that the amplitude of fluctuations is fixed on large scales by the CMB measurements, a low value of $\sigma_8$ can be caused by a non-zero neutrino mass which specifically reduces power on small scales, while leaving large scale power unchanged relative to standard $\Lambda$CDM.  The addition of these data sets yields a preferred mass of the sterile neutrino of around 0.5 eV, with $\Neff=4$ allowed.

\begin{table*}[ht]
\begin{center}
\resizebox{1\textwidth}{!}{
\begin{tabular}{|l|c|c|c|c|c|}
\hline
\hline
            &Planck+WP+high-$\ell$ 	&Planck+WP+high-$\ell$ 	&Planck+WP+high-$\ell$ 	&Planck+WP+high-$\ell$ &Planck+WP+high-$\ell$\\
Parameters  &+BICEP2			&+BICEP2		&+BICEP2		&+BICEP2		&+BICEP2\\
            &         			&+LSS         		&+$H_0$         	&+LSS+$H_0$ 		&+LSS+$H_0$+CFHTLenS+PSZ\\
\hline
$\Omega_{\rm b} h^2$ 	
			& $0.02251^{+0.00039}_{-0.00046}\,^{+0.00087}_{-0.00078}$ & $0.02232^{+0.00033}_{-0.00039}\,^{+0.00073}_{-0.00069}$ & $0.02257^{+0.00029}_{-0.00030}\,^{+0.00059}_{-0.00057}$ & $0.02248^{+0.00029}_{-0.00029}\,^{+0.00057}_{-0.00056}$ & $0.02267^{+0.00027}_{-0.00028}\,^{+0.00055}_{-0.00053}$ \\

$\Omega_{\rm cdm} h^2$ 
			& $0.129^{+0.006}_{-0.007}\,^{+0.013}_{-0.012}$           & $0.128^{+0.005}_{-0.006}\,^{+0.011}_{-0.010}$           & $0.130^{+0.006}_{-0.006}\,^{+0.011}_{-0.011}$           & $0.129^{+0.005}_{-0.005}\,^{+0.011}_{-0.011}$           & $0.127^{+0.006}_{-0.006}\,^{+0.011}_{-0.011}$           \\

$\theta_{\rm s}$ 
			& $1.0399^{+0.0009}_{-0.0009}\,^{+0.0017}_{-0.0017}$      & $1.0401^{+0.0009}_{-0.0008}\,^{+0.0017}_{-0.0017}$      & $1.0398^{+0.0008}_{-0.0008}\,^{+0.0018}_{-0.0016}$      & $1.0399^{+0.0008}_{-0.0008}\,^{+0.0017}_{-0.0016}$      & $1.0400^{+0.0009}_{-0.0009}\,^{+0.0018}_{-0.0017}$      \\

$\tau$ 
			& $0.097^{+0.013}_{-0.016}\,^{+0.031}_{-0.027}$           & $0.093^{+0.013}_{-0.014}\,^{+0.027}_{-0.027}$           & $0.099^{+0.013}_{-0.015}\,^{+0.029}_{-0.026}$           & $0.095^{+0.013}_{-0.014}\,^{+0.028}_{-0.027}$           & $0.091^{+0.013}_{-0.015}\,^{+0.028}_{-0.027}$           \\

$n_{\rm s}$ 
			& $0.986^{+0.016}_{-0.020}\,^{+0.035}_{-0.033}$           & $0.977^{+0.012}_{-0.016}\,^{+0.028}_{-0.027}$           & $0.989^{+0.011}_{-0.011}\,^{+0.021}_{-0.022}$           & $0.985^{+0.011}_{-0.010}\,^{+0.020}_{-0.022}$           & $0.993^{+0.010}_{-0.011}\,^{+0.021}_{-0.021}$           \\

$\log(10^{10} A_s)$ 
			& $3.120^{+0.030}_{-0.037}\,^{+0.071}_{-0.061}$           & $3.182^{+0.042}_{-0.038}\,^{+0.073}_{-0.078}$           & $3.124^{+0.030}_{-0.031}\,^{+0.060}_{-0.058}$           & $3.116^{+0.029}_{-0.030}\,^{+0.060}_{-0.055}$           & $3.124^{+0.031}_{-0.031}\,^{+0.063}_{-0.061}$           \\

$r$ 
			& $0.177^{+0.036}_{-0.050}\,^{+0.093}_{-0.086}$           & $0.168^{+0.034}_{-0.046}\,^{+0.085}_{-0.078}$           & $0.181^{+0.037}_{-0.047}\,^{+0.087}_{-0.081}$           & $0.175^{+0.035}_{-0.045}\,^{+0.083}_{-0.077}$           & $0.206^{+0.041}_{-0.051}\,^{+0.094}_{-0.090}$           \\
\hline
$\ns$ 
			& $0.82^{+0.40}_{-0.57};\,<1.66$                          & $0.61^{+0.25}_{-0.52};\,<1.30$                           & $0.88^{+0.32}_{-0.32}\,^{+0.64}_{-0.67}$                & $0.81^{+0.32}_{-0.32};\,<1.35$                          & $0.89^{+0.34}_{-0.37}\,^{+0.70}_{-0.69}$                \\

$m_s [\rm{eV}]$
			& $<0.85$                                                 & $<0.82$                                                 & $<0.50$                                                 & $<0.48$                                                 & $0.44^{+0.11}_{-0.16}\,^{+0.33}_{-0.32}$                \\

\hline
\hline
\end{tabular}
}
\vspace{0.3cm}
\caption{Marginalized $1\sigma$ and $2\sigma$ confidence level limits for the cosmological parameters in various dataset combinations, given with respect to the mean value. Upper limit are given at $2\sigma$.}
\label{tab:cosmoresults}
\end{center}
\end{table*}

{\it Adding SBL data ---}

The next question is how compatible the cosmological and SBL data really are. When we use cosmological data with lensing and cluster data excluded we find a relatively stringent upper bound on $m_s$. This is relaxed when $\ns$ is low, simply because the suppression of structure formation scales with the total density in neutrinos at late times, i.e.\ as $\ns^{3/4} m_s$. However, since CMB data prefers a high $\ns$ this possibility is disfavoured, and the conclusion is that CMB and LSS data requires the sterile mass to be low. Again, the bound can easily be relaxed in models where additional dark radiation is provided by other particles, but in the simple model discussed here it is disfavoured.
When we add lensing and cluster data the sterile mass comes out around 0.5 eV and with fully thermalised sterile neutrinos being allowed. 

In Tab.~\ref{tab:sblresults} we report the marginalized mean values and the $1\sigma$ and $2\sigma$ errors on the cosmological parameters and on the neutrino parameters in the different combinations of data sets illustrated above, when SBL data are included.

As we stated before, it is easy to see that the anti-correlation between $m_s$ and $\ns$, together with the strong bounds on $m_s$ from the SBL data, leaves a very small space to a fully thermalized sterile neutrino. When adding SBL data, the constraints on $m_s$ come only by the oscillation experiments, with very small dependence on the cosmological data. 
On the other hand, cosmology provides a strong limit on $\ns$ that is compatible with 0 within $2\sigma$ in all the cases that do not include CFHTLenS and PSZ data. When LSS data are included, the value of $\ns$ is more strongly constrained. Only when CFHTLenS and PSZ are included there is a little evidence that $\ns>0$ at more than $1\sigma$: even in this case, however, a fully thermalized sterile neutrino with $\ns=1$ is strongly disfavoured.

This tension between cosmology and SBL data, yet studied in past works (see e.g. \cite{Mirizzi:2013kva}) is not alleviated in the physical case (i.e. by allowing $\ns$ to vary in the range $[0,1]$): the mass values preferred by SBL data lay above the hot dark matter limit and therefore they are disfavoured by cosmology, even if there is only one partially (or fully $\ns=1$) thermalized sterile neutrino. Quantitatively speaking, a model with one fully thermalized sterile neutrino and with a mass fixed at the SBL best-fit ($m_s=1.27$ eV) compared to the cosmological best-fit model has a $\Delta \chi^2 \simeq 18$ if Planck+WP+high-$\ell$ data are considered. If also BICEP2 data are considered, the value lowers to $\Delta \chi^2 \simeq 12$: this is possible since the inclusion of the BICEP2 data strengthens the limit on the mass, but it weakens the limit on $\ns$. 

If a partial thermalization is taken into account and $\ns$ is free to vary moving towards lower values, the $\Delta\chi^2$ with respect to the best fit is lower. For a $m_s=1.27$ eV neutrino with small $\ns$ we have $\Delta \chi^2 \simeq 1$ from Planck+WP+high-$\ell$ and $\Delta \chi^2 \simeq 6$ from Planck+WP+high-$\ell$+BICEP2.

We can conclude that a fully thermalized sterile neutrino with a mass fixed at the SBL best-fit is less disfavoured by cosmology if the BICEP2 data are included. On the contrary if the sterile neutrino is not fully thermalized the inclusion of BICEP2 data worsens the consistency of this hypothesis with cosmology.

\begin{table*}[ht]
\begin{center}
\resizebox{1\textwidth}{!}{
\begin{tabular}{|l|c|c|c|c|c|}
\hline
\hline
&SBL+Planck+WP &SBL+Planck+WP &SBL+Planck+WP &SBL+Planck+WP &SBL+Planck+WP\\
 Parameters &+high-$\ell$+BICEP2 &+high-$\ell$+BICEP2 &+high-$\ell$+BICEP2 &+high-$\ell$+BICEP2 &+high-$\ell$+BICEP2\\
                   &         &+LSS         &+$H_0$          &+LSS+$H_0$ &+LSS+$H_0$+CFHTLenS+PSZ\\
\hline

$\Omega_{\rm b} h^2$ 	
			& $0.02214^{+0.00029}_{-0.00029}\,^{+0.00058}_{-0.00058}$ & $0.02200^{+0.00026}_{-0.00025}\,^{+0.00051}_{-0.00052}$ & $0.02230^{+0.00027}_{-0.00027}\,^{+0.00060}_{-0.00054}$ & $0.02214^{+0.00025}_{-0.00025}\,^{+0.00049}_{-0.00051}$ & $0.02236^{+0.00023}_{-0.00023}\,^{+0.00047}_{-0.00047}$ \\

$\Omega_{\rm cdm} h^2$ 
			& $0.121^{+0.003}_{-0.004}\,^{+0.008}_{-0.007}$           & $0.121^{+0.002}_{-0.003}\,^{+0.006}_{-0.005}$           & $0.118^{+0.003}_{-0.004}\,^{+0.007}_{-0.006}$           & $0.118^{+0.002}_{-0.002}\,^{+0.005}_{-0.005}$           & $0.117^{+0.002}_{-0.003}\,^{+0.006}_{-0.006}$           \\

$\theta_{\rm s}$ 
			& $1.0408^{+0.0008}_{-0.0007}\,^{+0.0015}_{-0.0014}$      & $1.0409^{+0.0006}_{-0.0006}\,^{+0.0012}_{-0.0013}$      & $1.0413^{+0.0007}_{-0.0006}\,^{+0.0013}_{-0.0015}$      & $1.0413^{+0.0006}_{-0.0006}\,^{+0.0012}_{-0.0012}$      & $1.0413^{+0.0006}_{-0.0006}\,^{+0.0013}_{-0.0014}$      \\

$\tau$ 
			& $0.092^{+0.012}_{-0.014}\,^{+0.026}_{-0.025}$           & $0.088^{+0.012}_{-0.014}\,^{+0.027}_{-0.024}$           & $0.094^{+0.012}_{-0.015}\,^{+0.028}_{-0.027}$           & $0.091^{+0.012}_{-0.014}\,^{+0.026}_{-0.024}$           & $0.086^{+0.012}_{-0.014}\,^{+0.026}_{-0.024}$           \\

$n_{\rm s}$ 
			& $0.962^{+0.008}_{-0.008}\,^{+0.016}_{-0.015}$           & $0.958^{+0.006}_{-0.006}\,^{+0.013}_{-0.013}$           & $0.967^{+0.007}_{-0.008}\,^{+0.015}_{-0.014}$           & $0.962^{+0.006}_{-0.006}\,^{+0.012}_{-0.012}$           & $0.970^{+0.005}_{-0.005}\,^{+0.011}_{-0.011}$           \\

$\log(10^{10} A_s)$ 
			& $3.213^{+0.031}_{-0.031}\,^{+0.063}_{-0.063}$           & $3.220^{+0.030}_{-0.030}\,^{+0.059}_{-0.059}$           & $3.091^{+0.026}_{-0.030}\,^{+0.057}_{-0.051}$           & $3.085^{+0.025}_{-0.027}\,^{+0.052}_{-0.048}$           & $3.169^{+0.027}_{-0.026}\,^{+0.053}_{-0.052}$           \\

$r$ 
			& $0.160^{+0.034}_{-0.042}\,^{+0.078}_{-0.075}$           & $0.150^{+0.032}_{-0.039}\,^{+0.071}_{-0.067}$           & $0.164^{+0.032}_{-0.043}\,^{+0.079}_{-0.073}$           & $0.158^{+0.032}_{-0.042}\,^{+0.075}_{-0.070}$           & $0.179^{+0.034}_{-0.043}\,^{+0.082}_{-0.076}$           \\
\hline
$\ns$ 
			& $<0.63$                                                 & $<0.28$                                                 & $<0.59$                                                 & $<0.22$                                                 & $0.19^{+0.07}_{-0.15};\,<0.42$                          \\

$m_s [\rm{eV}]$
			& $1.21^{+0.14}_{-0.13}\,^{+0.19}_{-0.25}$                & $1.22^{+0.13}_{-0.13}\,^{+0.20}_{-0.25}$                & $1.20^{+0.14}_{-0.12}\,^{+0.19}_{-0.25}$                & $1.21^{+0.14}_{-0.13}\,^{+0.19}_{-0.26}$                & $1.19^{+0.15}_{-0.12}\,^{+0.19}_{-0.25}$                \\

\hline
\hline
\end{tabular}
}
\vspace{0.3cm}
\caption{Marginalized $1\sigma$ and $2\sigma$ confidence level limits for the cosmological parameters in various dataset combinations, given with respect to the mean value. Upper limit are given at $2\sigma$.}
\label{tab:sblresults}
\end{center}
\end{table*}

\section{Discussion}

We have performed an analysis of light sterile neutrinos in the context of both cosmology and short baseline neutrino oscillation experiments. Previous analyses have shown that while SBL data points to the existence of a mainly sterile mass state around 1 eV, this is not compatible with cosmological data unless the additional state is somehow prevented from being fully thermalised in the early Universe \cite{Archidiacono:2012ri}.

The inclusion of new data from the BICEP-2 experiment favours a higher dark radiation content, but this actually tightens the cosmological bound on the mass of the sterile neutrino because $m_s$ and $\ns$ are highly anti-correlated.
Cosmological data from the CFHTLenS survey and the Planck SZ cluster counts actually favour a non-zero mass of the sterile neutrino because it alleviates the tension between the value of $\sigma_8$ inferred from the CMB measurements and the minimal $\Lambda$CDM model and the lower value indicated by data CFHTLenS and PSZ data. The inclusion of these two data sets points to a sterile mass around 0.5 eV, but with relatively low $\ns$. Provided that $\ns$ is low the allowed mass stretches to higher values.

The SBL data strongly constrains $m_s$, but not $\ns$, and indicates a mass not much lower than 1 eV. At the same time the mixing angle is large enough that the additional state is almost fully thermalised. However, this scenario is highly disfavoured by cosmological data (with a $\Delta \chi^2>10$) which for a mass of 1 eV requires $\ns$ to be quite low. Indeed a model with a mass of 1 eV and a low $\ns$ is compatible with cosmology within roughly $2\sigma$ confidence level.
The conclusion is that light sterile neutrinos as indicated by SBL data are close to being ruled out by cosmological data unless they are somehow prevented from thermalising in the early Universe. 
Recently, a number of papers on how to resolve this apparent conflict have appeared.

A possible way out of this problem is that sterile neutrinos have new interactions which induce a non-standard matter potential and block thermalisation \cite{Hannestad:2013ana,Dasgupta:2013zpn,Bringmann:2013vra,Ko:2014bka,Mirizzi:2012we,Saviano:2013ktj}. This model can easily have 1 eV sterile neutrinos and an $\Neff$ not much beyond 3 and thus be compatible with all existing data.
While this scenario certainly works well and can possibly also explain some of the astrophysical anomalies related to cold dark matter, there are without a doubt other possible ways of making eV sterile neutrino compatible with both SBL and cosmological data. For example some models with low temperature reheating or non-standard expansion rate of the universe at the MeV scale where the new state is thermalised can also prevent thermalisation \cite{Rehagen:2014vna}.
Thus, eV mass sterile neutrinos remain an intriguing possibility which potentially has wide ranging implications for cosmology.

\section*{Acknowledgments}
MA acknowledges European ITN project Invisibles (FP7-PEOPLE-2011-ITN, PITN-GA-2011-289442-INVISIBLES).
This
work is supported by the research grant {\sl Theoretical Astroparticle Physics} number 2012CPPYP7 under the program PRIN 2012  funded by the Ministero dell'Istruzione, Universit\`a e della Ricerca (MIUR), by the research grant {\sl TAsP (Theoretical Astroparticle Physics)}
funded by the Istituto Nazionale di Fisica Nucleare (INFN), by the  {\sl Strategic Research Grant: Origin and Detection of Galactic and Extragalactic Cosmic Rays} funded by Torino University and Compagnia di San Paolo, by the Spanish MINECO under grants FPA2011-22975 and MULTIDARK CSD2009-00064 (Consolider-Ingenio 2010 Programme).

\bibliographystyle{utcaps}

%\bibliography{refs}

\providecommand{\href}[2]{#2}\begingroup\raggedright\endgroup

\end{document}